\documentstyle[11pt,epsfig,amssymb,theorem]{article}
\newcommand{\be}{\begin{equation}}
\newcommand{\ee}{\end{equation}}
\newcommand{\ba}{\begin{eqnarray}}
\newcommand{\ea}{\end{eqnarray}}
\newcommand{\E}{\rm E}

\theoremstyle{break}
\newtheorem{theorem}{Theorem}
\newtheorem{proposition}{Proposition}

\newtheorem{definition}{Definition}

[1999/03/29 PSNFSS v.7.2
Times font as default roman
: S Rahtz]

\begin{document}

\title{Multi-dimensional Rational Bubbles and fat tails
\footnote{We acknowledge helpful discussions and exchanges with J.P.
Laurent,
T. Lux, V. Pisarenko and M. Taqqu and thank T. Mikosch for providing
access
to \cite{LP83}.}
}

\author{Y. Malevergne$^1$ and D. Sornette$^{1,2}$ \\
$^1$ Laboratoire de Physique de la Mati\`ere Condens\'ee\\
CNRS UMR 6622 and
Universit\'e de Nice-Sophia Antipolis\\ 06108 Nice Cedex 2, France\\
$^2$ Institute of Geophysics and Planetary Physics\\
and Department of Earth and Space Science\\
University of California, Los Angeles, California 90095, USA
}

\maketitle

\begin{abstract}

We extend the model of rational bubbles of
\cite{Blanchard1} and \cite{Blanwat} to arbitrary dimensions $d$:
a number $d$ of
market time series are made linearly interdependent via $d \times d$
stochastic coupling
coefficients. We first show that the no-arbitrage
condition imposes that the non-diagonal impacts of any asset $i$ on any
other
asset $j \neq i$ has to vanish on average, i.e., must exhibit
random alternative regimes of reinforcement and contrarian feedbacks.
In contrast, the diagonal terms must be positive and equal on average
to the inverse
of the discount factor. Applying the results of renewal theory for
products
of random matrices to stochastic recurrence equations (SRE), we extend the
theorem of \cite{Luxsor} and
demonstrate that the tails of the unconditional distributions
associated with such
$d$-dimensional bubble processes follow power laws
(i.e., exhibit hyperbolic decline), with the
same asymptotic tail exponent $\mu<1$ for all assets. The
distribution of price differences and of
returns is dominated by the same power-law over an extended range
of large returns. This small value  $\mu<1$ of the tail exponent
has far-reaching consequences in the non-existence of the means and
variances.
Although power-law tails are a pervasive feature of empirical
data, the numerical value  $\mu<1$ is in disagreement with the usual
empirical
estimates $\mu \approx 3$. It, therefore, appears that generalizing
the model of rational bubbles to arbitrary dimensions does not allow us
to reconcile the model with these stylized facts of financial data. The
non-stationary growth rational bubble model seems at present the only
viable solution
\cite{growthbubble}.

\end{abstract}

\section{The model of rational bubbles}

\cite{Blanchard1} and \cite{Blanwat} originally introduced the model of
rational expectations (RE) bubbles to account for the possibility, often
discussed in the empirical literature and by practitioners, that observed
prices may deviate significantly and over extended time intervals
from fundamental prices.
While allowing for deviations from fundamental prices, rational bubbles
keep
a fundamental anchor point of economic modelling, namely that bubbles
must obey the condition of rational expectations. In contrast,
recent works stress
that investors are not fully rational, or have at most bound
rationality, and that behavioral and psychological
mechanisms, such as herding, may be important in the shaping of market
prices
\cite{Thaler,Shefrin,Shleifer}.
However, for fluid assets, dynamic investment strategies rarely perform
over
simple buy-and-hold strategies
\cite{Malkiel}, in other words, the market is not far from being efficient
and little arbitrage opportunities exist as a result of the constant
search for gains by sophisticated investors. Here, we shall work within
the conditions of rational expectations and of no-arbitrage condition,
taken as
useful approximations. Indeed, the rationality of both expectations
and behavior often
does not imply that the price of an asset be equal to its fundamental
value. In other words, there can be rational deviations of the price from
this value, called rational bubbles. A rational bubble can arise when
the actual
market price depends positively on its own expected rate of change, as
sometimes occurs in asset markets, which is the mechanism underlying the
models
of \cite{Blanchard1} and \cite{Blanwat}.

In order to avoid the
unrealistic picture of ever-increasing deviations from fundamental values,
\cite{Blanwat} proposed a model with periodically collapsing bubbles in
which
the bubble component of the price follows an
exponential explosive path (the price being multiplied
by $a_t={\bar a}>1$) with probability $\pi$ and collapses to zero
(the price being multiplied
by $a_t=0$) with probability $1 - \pi$. It is clear that, in this model,
a bubble has an exponential
distribution of lifetimes with a finite average lifetime $\pi/(1-\pi)$.
Bubbles are thus transient phenomena. The condition
of rational expectations imposes that ${\bar a}=1/\delta$, where
$\delta$ is the
inverse of the discount factor.
In order to allow for the start of new bubbles after the collapse, a
stochastic
zero mean normally distributed component $b_t$ is added to the
systematic part of $B_t$.
This leads to the following dynamical equation
\be
X_{t+1} = a_t X_t + b_t,    \label{eq1}
\ee
where, as we said,
   $a_t={\bar a}$ with probability $\pi$ and $a_t=0$ with probability $1 -
\pi$.
There is a huge literature
on theoretical refinements of this model and on the empirical
detectability of RE bubbles in financial data (see
\cite{Camerer} and \cite{adamsz}, for surveys of this literature).

Recently, \cite{Luxsor} studied the implications of the RE bubble models
for
the unconditional distribution of prices, price
changes and returns resulting from a more general discrete-time
formulation
extending (\ref{eq1}) by allowing the multiplicative factor
$a_t$ to take arbitrary values and be
i.i.d. random variables drawn from some non-degenerate
probability density function (pdf)
$P_a(a)$. The model can also be generalized by considering
non-normal realizations of $b_t$ with distribution
$P_b(b)$ with $\E[b_t]=0$, where $\E[\cdot]$ is the expectation operator.
Since in (\ref{eq1}) the bubble $X_t$ denotes the difference
between the observed price and the fundamental price,
the ``bubble'' regimes refer to the cases when $X_t$ explodes
exponentially
under the action of successive multiplications by factor $a_t, a_{t+1},
...$
with a majority of them larger than $1$ but different, thus adding
a stochastic component to the standard model of
\cite{Blanwat}.

Provided $\E[\ln a] < 0$ (stationarity
condition) and if there is a number $\mu$ such that
$0 < \E[|b|^{\mu}] < +\infty$, such that
\be
\E[|a|^{\mu}] = 1    \label{nfakka}
\ee
and such that
$\E[|a|^{\mu} \ln |a|] < +\infty$, then the tail of the distribution of
$B$ is asymptotically (for large $X$'s) a power law \cite{K73,Goldie}
\be
P_X(X) ~dX \approx {C  \over |X|^{1+\mu}}~dX~,   \label{fkaka}
\ee
with an exponent $\mu$ given by the real positive solution of
(\ref{nfakka}).
Rational expectations require in addition that
the bubble component of asset prices obeys the ``no free-lunch'' condition
\be
\delta \cdot \E[X_{t+1}] = X_t    \label{bjqjak}
\ee
where $\delta$ is the discount factor $< 1$. Condition (\ref{bjqjak})
with (\ref{eq1}) imposes the condition
\be
\E[a] = 1/\delta  >1~,
\ee
on the distribution of the multiplicative factors $a_t$.
Since the function $\E[|a|^{\mu}]$ is convex, \cite{Luxsor}
showed that this automatically enforces $\mu < 1$.
It is easy to show that the distribution of price differences has the same
power law tail
with the exponent $\mu<1$ and the distribution of returns is dominated by
the same power-law over an extended range of large returns \cite{Luxsor}.
Although power-law
tails are a pervasive feature of empirical data, these characterizations
are in strong disagreement with the usual empirical estimates which
find $\mu \approx 3$
\cite{devries,Lux,Pagan,Guillaume1,Gopikrishnan}.
\cite{Luxsor} concluded that
exogenous rational bubbles are thus hardly reconcilable with some of
the stylized
facts of financial data at a very elementary level.

\section{Generalization of rational bubbles to arbitrary dimensions}

\subsection{Introduction}

In reality, there is no such thing as an isolated asset. Stock markets
exhibit
a variety of inter-dependences, based in
part on the mutual influences between the USA, European and
Japanese markets. In addition, individual stocks may be sensitive to
the behavior of the specific industry as a whole to which they belong
and to a few other
indicators, such as the main indices, interest rates and so on.
\cite{Mantegna1,Mantegna2}
have indeed shown the existence of a
hierarchical organization of stock interdepences. Furthermore,
bubbles often appear to be not isolated features of a set of markets.
For instance, \cite{Flood} tested whether a
bubble simultaneously existed across the nations, such as
Germany, Poland, and Hungary, that experienced
hyperinflation in the early 1920s. Coordinated bubbles can sometimes
be detected. One of the most prominent example is found in the market
appreciations
observes in many of the world markets prior to the world market crash
in Oct. 1987 \cite{presidentcomit}.
Similar intermittent coordination of bubbles
have been detected among the significant bubbles followed
by large crashes or severe corrections in
Latin-American and Asian stock markets \cite{emer}.
It is therefore desirable to generalize the one-dimensional
RE bubble model (\ref{eq1}) to the multi-dimensional case.
One could also hope a priori that this generalization would modify the
result $\mu<1$ obtained in the one-dimensional case and allow for a
better
adequation with empirical results. As we shall show, this turns out to be
ill-born for reasons that we shall explain in details.

The simplest such generalization is to consider two assets $X$ and
$Y$ with prices
respectively equal to $X_t$ and $Y_t$ at time $t$, evolving according to
\be
X_{t+1} = a_t X_t + b_t Y_t + \eta_t    \label{eq11}
\ee
\be
Y_{t+1} = c_t X_t + d_t Y_k + \epsilon_t    \label{eq12}
\ee
where $a_t$, $b_t$, $c_t$ and $d_t$ are drawn from some
multivariate probability density function. The two additive noises
$\eta_t$ and $\epsilon_t$ are also drawn from some distribution function
with zero mean.
The diagonal case $b_t=c_t=0$ for all $t$ recovers the previous
one-dimensional case with two uncoupled bubbles, provided $\eta_t$ and
$\epsilon_t$ are independent.

Rational expectations require that $X_t$ and $Y_t$
obey both the ``no-free lunch'' condition (\ref{bjqjak}),
i.e., $\delta \cdot \E[X_{t+1}] = X_t$ and $\delta \cdot \E[Y_{t+1}] =
Y_t$.
With (\ref{eq11},\ref{eq12}), this gives
\ba
\left(\E[a_t]-\delta^{-1}\right) X_t + \E[b_t] Y_t &=& 0~,   \label{hgaal}
\\
\E[c_t]X_t + \left(\E[d_t]-\delta^{-1}\right)  Y_t &=& 0~,  \label{fjkala}
\ea
where we have used $\E[\eta_t]=\E[\epsilon_t]=0$. The two equations
(\ref{hgaal},\ref{fjkala}) must be true for all times, i.e. for all
values of $X_t$ and $Y_t$ visited by the dynamics. This imposes
$\E[b_t]=\E[c_t]=0$ and $\E[a_t]=\E[d_t]=\delta^{-1}$.
We are going to retrieve this result more formally in the general case.

\subsection{General formulation}

A generalization to arbitrary dimensions lead to the
following stochastic random equation (SRE)
\be
\label{eq:sre}
{\bf X_t} = {\bf A_t X_{t-1}} + {\bf B_t}
\ee
where $({\bf X_t},{\bf B_t})$ are d-dimensional vectors. Each
component of ${\bf X_t}$
can be thought of as the price of an asset above its fundamental price.
The matrices $({\bf A_t})$ are
identically independent distributed $d \times d$-dimensional
stochastic matrices. We assume that ${\bf B_t}$ are
identically independent distributed random vectors and that $({\bf
X_t})$ is a causal
stationnary solution of (\ref{eq:sre}). Generalizations introducing
additional arbitrary linear
terms at larger time lags such as $X_{t-2}, ...$ can be treated
with slight modifications of our approach and yield the same
conclusions. We shall
thus confine our demonstration on the SRE of order $1$, keeping in mind
that
our results apply analogously to arbitrary orders of regressions.

To formalize the SRE in a rigorous manner, we
introduce in a standard way the probability space
$(\Omega, {\cal F}, {\mathbb P})$ and a filtration $({\cal F}_t)$.
Here ${\mathbb P}$ represents the product measure ${\mathbb P}={\mathbb
P}_X
\otimes {\mathbb P}_A \otimes {\mathbb P}_B$, where ${\mathbb P}_X$,
${\mathbb P}_A$ and
${\mathbb P}_B$, are the probability measures associated with $\{{\bf
X_t}\}$,
$\{{\bf A_t}\}$ and $\{{\bf B_t}\}$.
We further assume as is customory that the stochastic
process $({\bf X_t})$ is adapted to the filtration $({\cal F}_t)$.

In the following, we denote by $|\cdot|$ the Euclidean norm and by
$||\cdot||$
the corresponding norm for any
$d \times d$-matrix ${\bf A}$
\be
||{\bf A}|| = \sup_{|{\bf x}|=1} |{\bf Ax}|  .
\ee

In the next section \ref{sec:nfl}, we formalize the ``no-free lunch''
condition for the SRE
(\ref{eq:sre}) and show that its entails in particular that the spectral
radius
(larger eigenvalue) of $\E[{\bf A}_t]$ must be equal to the inverse of the
discount factor,
hence it must be larger than $1$. In section \ref{sectrenkes}, we
synthetize the
main results on the renewal theory of SRE of the type (\ref{eq:sre}) and
show that the condition imposing that the exponent of the power law
tails be larger
than $1$ implies that the spectral radius of $\E[{\bf A}_t]$ must be
smaller than $1$.
By the reverse logic, the ``no-free lunch'' condition automatically
enforces our main
result that the exponent is less than $1$.

\section{The no-free lunch condition}
\label{sec:nfl}
\subsection{No-free lunch condition under the risk-neutral probability
measure}

The ``no-free lunch'' condition is equivalent to the existence of a
probability measure
${\mathbb Q}$ equivalent to ${\mathbb P}$ such that, for all
self-financing portfolio
$P_t$, ${ P_t\over S_{0,t}}$ is a
${\mathbb Q}$-martingale, where $S_{0,t}=\prod_{i=0}^{t-1}
{\delta_i}^{-1}$,
$\delta_i=(1+r_i)^{-1}$ is the discount factor for period $i$ and
$r_i$ is the corresponding risk-free interest rate.

It is natural to assume that, for a given period $i$, the discount rates
$r_i$ are the same for all assets. In frictionless markets, a
deviation for this
hypothesis would lead to arbitrage opportunities.
Furthermore, since the sequence of matrices $\{ \bf{A}_t \}$ is i.i.d.
and therefore stationnary,
this implies that $\delta_t$ or $r_t$ must be constant and equal
respectively to $\delta$
and $r$.

Under those conditions, we have the following proposition :

\begin{proposition}
The stochastic process
\be
{\bf X_t} = {\bf A_t X_{t-1}} + {\bf B_t}
\ee
satisfies the no-arbitrage condition {\it if and only if}
\be
E_{\mathbb Q}[{\bf A}] = \frac{1}{\delta} {\bf I_d}~.
\label{condnoarbi}
\ee
\end{proposition}

The proof is given in the Appendix A.

The condition (\ref{condnoarbi}) imposes some stringent constraints on
admissible matrices ${\bf A_t}$. Indeed, while ${\bf A_t}$ are not
diagonal in general,
their average must be diagonal. This implies that the off-diagonal terms
of
the matrices ${\bf A_t}$ must take negative values, sufficiently often so
that their averages vanish. The off-diagonal coefficients quantify
the influence
of other bubbles on a given one. The condition (\ref{condnoarbi})
thus means that
the average effect of other bubbles on any given one must vanish. It is
straightforward to check that, in this linear framework, this implies an
absence of correlation between the different bubble components
$\E[X^{(k)} X^{(\ell)}]=0$ for any $k \neq \ell$.

In constrast, the diagonal elements of  ${\bf A_t}$ must be positive
in majority in order for $E_{\mathbb P}[A_{ii}]={\delta^{(i)}}^{-1}$,
for all $i$'s, to hold true.
In fact, on economic grounds, we can exclude the cases
where the diagonal elements take negative values. Indeed, a negative value
of
$A_{ii}$ at a given time $t$ would imply that $X_t^{(i)}$ abruptly
change sign between
$t-1$ and $t$, what does not seem to be a reasonable financial process.

\subsection{The no-free lunch condition under historic probability
measure}
\label{sec:nflrs}

The historical ${\mathbb P}$ and risk-neutral ${\mathbb Q}$
probability measures are equivalent. This means that there exists a
non-negative matrix ${\bf h}({\bf \theta}) = \left(h_{ij}(\theta_{ij})\right)$ such
that,
for each element indexed by $i,j$, we have
\ba
E_{\mathbb P}[A_{ij}] &=& E_{\mathbb Q}[h_{ij}\cdot A_{ij}]\\
&=& h_{ij}(\theta_{ij}^0)\cdot E_Q[ A_{ij}] ~ ~\mbox{for}~\mbox{some}~
\theta_{ij}^0 \in {\mathbb R}~.
\ea
The second equation comes from the well known result~:
\be
\int f(\theta) \cdot g(\theta) ~ d\mu(\theta) = g(\theta_0) \cdot
\int f(\theta) ~d\mu(\theta) ~
~\mbox{for}~\mbox{some}~\theta_0 \in {\mathbb R}~.
\ee

We thus get
\ba
E_{\mathbb P}[A_{ij}] &=& 0~ ~\mbox{if}~ i\neq j\\
E_{\mathbb P}[A_{ij}] &=& \frac{1}{\delta^{(i)}}~ ~\mbox{if}~ i =j~,
\ea
where the $\delta^{(i)}$ can be different. We can thus write
\be
E_{\mathbb P}[{\bf A}]= \widetilde {\delta^{-1}}~,~~~
{\rm where}~~~\widetilde {\delta^{-1}} = \mbox{diag}[{\delta^{(1)}}^{-1},
\cdots ,{\delta^{(d)}}^{-1}]~.
\ee

Appendix B gives a proof showing that ${\delta^{(i)}}^{-1}$ is indeed
the genuine
discount factor for the $i$-th bubble component.

\section{Renewal theory for products of random matrices
\label{sectrenkes}}

In the following, we will consider that the random
$d \times d$ matrices ${\bf A}_t$ are invertible matrices with real
entries.
We will denote by $GL_d(\mathbb{R})$ the group of these matrices.

\subsection{Definitions}

\begin{definition}[Feasible matrix]
A matrix $M \in GL_d(\mathbb{R})$ is ${\mathbb P}$-feasible if there
exists an
$n \in \mathbb{N}$ and $M_1, \cdots, M_n \in$ supp$({\mathbb P})$ such
that
$M=M_1\cdots M_n$ and if $M$ has a  simple real eigenvalue $q(M)$ which,
in
modulus, exceeds all other  eigenvalue of $M$.
\end{definition}

\begin{definition}
For any matrix $M \in GL_d(\mathbb{R})$ and $M'$ its transpose, $MM'$
is a symmetric
positive definite matrix. We define $\lambda(M)$ the square root of
the smallest
eigenvalue of $MM'$.
\end{definition}

\subsection{Theorem}

We extend the theorem 2.7 of Davis et al. \cite{DMB99}, which
synthetized  Kesten's theorems 3 et 4 in \cite{K73}, to the case of
real valued matrices. The proof of this theorem is given in \cite{LP83}.

\begin{theorem}
Let $({\bf A}_n)$ be an i.i.d. sequence of matrices in
$GL_d(\mathbb{R})$ satisfying
the following set of conditions:
\begin{itemize}
\item[H1 :] for some $\epsilon > 0$, $E_{{\mathbb P}_A}\left[||{\bf
A}||^\epsilon
\right] <1$~,

\item[H2 :] For every open $U \subset S_{d-1}$ (the unit sphere in
${\mathbb R}^d$)  and for all $x \in S_{d-1}$
there exists an $n$ such that
\be
\Pr \left\{ \frac{{\bf x A_1 \cdots A_n}}{||{\bf x A_1 \cdots A_n}||}
\in U \right\} >0~.
\ee

\item[H3 :] The group $\{\ln |q(M)|, M~\mbox{is}~{\mathbb P}_A
\mbox{-feasible} \}$ is dense in $\mathbb{R}$~.

\item[H4 :] for all ${\bf r} \in {\mathbb R}^d$, $\Pr\{{\bf A_1 r} + {\bf
B_1}={\bf r}\}<1$~.

\item[H5 :] There exists a $\kappa_0>0$ such that
\be
E_{{\mathbb P}_A}\left(\left[ \lambda({\bf A_1})
\right]^{\kappa_0}\right) \ge 1~.
\ee

\item[H6 :] With the same $\kappa_0>0$ as for the previous condition,
there
exists
a real number $u >0$ such that
\be
\left\{
\begin{array}{l}
E_{{\mathbb P}_A} \left( \left[ \sup\{||{\bf A_1}||, ||{\bf B_1}|| \}
\right]^{\kappa_0+u}\right) < \infty~, \\
E_{{\mathbb P}_A}\left( ||{\bf A_1}||^{-u}\right) < \infty~.
\end{array}
\right.
\ee
\end{itemize}

\vspace{0.5cm}

Provided that these conditions hold,
\begin{itemize}
\item there exists a unique solution $\kappa_1 \in (0, \kappa_0]$ to the
equation \be
\label{eq:exp}
\lim_{n \rightarrow \infty} {1 \over n} \ln E_{{\mathbb P}_A}\left[||{\bf
A}_1
\cdots {\bf A}_n||^{\kappa_1}\right] = 0,
\ee
\item If $({\bf X}_n)$ is the stationary solution to the SRE in
(\ref{eq:sre})
then ${\bf X}$ is regularly varying with index $\kappa_1$. In other words,
the tail of the marginal distribution of each of the
components of the vector ${\bf X}$ is asymptotically a power law with
exponent $\kappa_1$.
\end{itemize}
\end{theorem}

\subsection{Comments on the theorem}

\subsubsection{Intuitive meaning of the hypotheses}

  A suitable property for an ecomonic model
is the stationnarity, i.e. the solution ${\bf X}_t$ of the SRE
(\ref{eq:sre})
does not blow up. This condition is ensured by the hypothesis ({\it
H1}). Indeed, $E_{{\mathbb P}_A} \left[\ln ||{\bf A}|| \right]<0$
implies that the Lyapunov exponent of the sequence $\{{\bf A}_n\}$
of i.i.d. matrices is negative \cite{DMB99}.
And it is well known, that the negativity of the Lyapunov exponent is a
sufficient condition for the existence of a stationnary solution ${\bf
X}_t$.

However, this condition can lead to a too fast decay of the tail of the
distibution of $\{{\bf X}\}$. This phenomenon is prevented by ({\it H5})
which
means intuitively that the multiplicative factors given by the
elements of ${\bf A}_t$ sometimes produce an amplification of ${\bf X_t}$.
In
the one-dimensional bubble case, this condition reduces to the simple
rule that $a_t$ must sometimes be larger than $1$ so that a bubble can
develop. Otherwise, the power law tail will be replaced by an exponential
tail.

So, ({\it H1}) and ({\it H5}) keep the balance between two opposite
objectives : to obtain a stationnary solution and to observe a fat tailed
distribution for the process (${\bf X_t}$).

Another desirable property for the model is the ergodicity~: we expect the
price process (${\bf X_t}$) to explore the entire space ${\mathbb R}^d$.
This is ensured by hypothesis ({\it H2}) and ({\it H4})~: hypothesis
({\it H2}) allows ${\bf X_t}$ to visit the neighborhood of any point in
${\mathbb R}^d$, and ({\it H4}) forbids the trajectory to be trapped
at some points.

Hypothesis ({\it H3}) and ({\it H6}) are more technical ones. The
hypothesis ({\it H6}) simply ensures that the tails of the distributions
of ${\bf A}_t$ and ${\bf B}_t$ are thinner than the tail created by the
SRE
(\ref{eq:sre}), so that the observed tail index is really due to the
dynamics
of the system and not to an ill-posed problem where the tail distribution
of ${\bf A}_t$ or  ${\bf B}_t$ is fat enought to dominate the large
deviations behavior of the process.
The hypothesis ({\it H3}) expresses some
kind of aperiodicity condition.

\subsubsection{Intuitive meaning of (\ref{eq:exp})}

The equation (\ref{eq:exp}) determining the tail exponent $\kappa_1$
reduces
to (\ref{nfakka}) in the one-dimensional case, which is simple to handle.
In the multi-dimensional case, the novel feature is the non-diagonal
nature of the multiplication of matrices which does not allow in general
for an explicit equation similar to (\ref{nfakka}).

\subsection{Constraint on the tail index}

The first conclusion of the theorem above shows that the tail index
$\kappa_1$ of the process $({\bf X_t})$ is driven by the behavior of the
matrices (${\bf A_t}$). We will then state a proposition in which we
give a majoration of the tail index.

\begin{proposition}
A necessary condition to have $\kappa_1 >1$ is that the spectral radius
of $E_{{\mathbb P}_A}[{\bf A}]$ be smaller than $1$~:
\be
\kappa_1 > 1 \Longrightarrow \rho(E_{{\mathbb P}_A}[{\bf A}]) <1~.
\ee
\end{proposition}

The proof is given in the Appendix C.

  This proposition, put together with
Proposition 1 above, will allow us to derive our main result.

\section{Consequences for rational expectation bubbles}

We have seen in section \ref{sec:nfl} from Proposition 1 that,
as a result of the no-arbitrage condition, the spectral radius of the
matrix $E_{{\mathbb P}} [{\bf A}]$ is greater than $1$. As a consequence,
by application of the converse of Proposition 2, we find that the
tail index $\kappa_1$ of the distribution of $({\bf X})$ is smaller
than $1$. This result does not rely on the diagonal property of the
matrices
$\E[{\bf A}_t]$ but only on the value of the spectral radius
imposed by the no-arbitrage condition.

This result generalizes to arbitrary $d$-dimensional processes the
result of \cite{Luxsor}. As a consequence, $d$-dimensional
rational expectation bubbles linking several assets suffer from the same
discrepancy compared to empirical data as the one-dimensional bubbles.
It would therefore appear that exogenous rational bubbles are hardly
reconcilable with some of the most
fundamental stylized facts of financial data at a very elementary level.

A possible remedy has recently been suggested \cite{growthbubble}
which involves an average exponential growth of the fundamental price at
some return rate $r_f>0$.
With the condition that the price fluctuations
associated with bubbles must on average grow with the mean market return
$r_f$,
it can be shown that the exponent of the power
law tail of the returns is no more bounded by $1$ as soon as $r_f$
is larger than the discount rate $r$ and can take essentially
arbitrary values. It would be interesting to investigate the interplay
between inter-dependence between several bubbles and this exponential
growth
model.

\pagebreak

\section*{APPENDIX A: proof of Proposition 1 on the no-arbitrage
condition}

Let $P_t$ be the value at time $t$ of any self-financing portfolio :
\be
P_t={\bf W_t}' {\bf X_t},
\ee
where ${\bf W_t}'=(W_1,....,W_d)$ is the vector whose componants are
the weight of
the different assets and the prime denotes the transpose.
The no-free lunch condition reads : \be
\label{eq:nfl}
P_t = \delta \cdot E_{\mathbb Q}[P_{t+1}|{\cal F}_t]~ ~ ~~~~~~~
\forall \{P_t\}_{t\ge 0}~.
\ee
Therefore, for all self-financing strategies  $({\bf W_t})$, we have
\be
{\bf W_{t+1}}' \left\{ E_{\mathbb Q}[{\bf A}] - \frac{1}{\delta} {\bf I_d}
\right\} {\bf X_t}=0 ~ ~ ~\forall~{\bf X_t} \in {\mathbb R}^d~,
\ee
where we have used the fact that
$({\bf W_{t+1}})$ is $({\cal F}_t)$-measurable and that the sequence of
matrices $\{ {\bf A_t} \}$ is i.i.d.

The strategy ${\bf W_t}'=(0,\cdots,0,1,0,\cdots,0)$ ($1$ in $i^{th}$
position), for
all $t$, is self-financing and implies
\be
(a_{i1}, a_{i2},\cdots, a_{ii}-\frac{1}{\delta},\cdots,a_{id}) \cdot
(X_t^{(1)},X_t^{(2)},\cdots,X_t^{(i)},\cdots, X_t^{(d)})'=0~,
\ee
for all ${\bf X_t} \in {\mathbb R}^d$. We have called $a_{ij}$ the
$(i,j)^{th}$ coefficient of the matrix $E_{\mathbb Q}[{\bf A}]$. As a
consequence,
\be
(a_{i1}, a_{i2},\cdots, a_{ii}-\frac{1}{\delta},\cdots,a_{id})=0 ~
~~~~~\forall i~,
\ee
and
\be
E_{\mathbb Q}[{\bf A}] = \frac{1}{\delta} {\bf I_d}~.
\ee
The no-arbitrage contidition thus implies : $E_{\mathbb Q}[{\bf A}] =
\frac{1}{\delta} {\bf I_d}$.

We now show that the converse is true, namely that
if $E_{\mathbb Q}[{\bf A}] = \frac{1}{\delta} {\bf I_d}$ is true, then
the no-arbitrage condition is verified.
Let us thus assume that $E_{\mathbb Q}[{\bf A}] = \frac{1}{\delta}
{\bf I_d}$ hold.
Then
\ba
E_{\mathbb Q}[P_{t+1}|{\cal F}_t] &=& E_{\mathbb Q}[{\bf W_{t+1}}' \cdot
{\bf X_{t+1}}|{\cal F}_t]\\
&=& {\bf W_{t+1}}' \cdot E_{\mathbb Q}[{\bf X_{t+1}}|{\cal F}_t]\\
&=& {\bf W_{t+1}}' \cdot E_{\mathbb Q}[{\bf A_{t+1} X_t} + {\bf
B_{t+1}}|{\cal F}_t]\\
&=& {\bf W_{t+1}}' \cdot E_{\mathbb Q}[{\bf A_{t+1}}|{\cal F}_t]
\cdot {\bf X_t}\\
&=& {\bf W_{t+1}}' \cdot E_{\mathbb Q}[{\bf A}] \cdot {\bf X_t}\\
&=& \frac{1}{\delta}{\bf W_{t+1}}' \cdot {\bf X_t}~.
\ea

The condition that the portfolio is self-financing is ${\bf W_{t+1}'
X_t}={\bf W_t' X_t}$,
which means that the weights can be rebalanced a priori arbitrarily
between the assets
with the constraint that the total wealth at the same time remains
constant.
We can thus write
thus
\ba
E_{\mathbb Q}[P_{t+1}|{\cal F}_t] &=& \frac{1}{\delta}{\bf W_{t+1}}'
\cdot {\bf X_t}\\
&=& \frac{1}{\delta}P_t~.
\ea
Therefore, the discounted process $\{P_t\}$ is a ${\mathbb Q}$-martingale.

\section*{APPENDIX B: proof that ${\delta^{(i)}}^{-1}$ is the
discount factor for the $i$-th bubble component in the historical space}

Here, we express the no-free lunch condition in the historical space (or
real space). The condition we will obtain is the so-called ``Rational
Expectation Condition'', which is a little bit less general than the
condition
detailed in the previous appendix A.

Given the prices $\{X_k^{(i)}\}_{k \le t}$ of an asset, labeled by
$i$, until the date $t$,
the best estimation of its price at $t+1$ is $E_{\mathbb
P}[X_{t+1}^{(i)}| {\cal F}_{t}]$.
So, the RE condition leads to
\be
\frac{E_{\mathbb P}[X_{t+1}^{(i)}| {\cal
F}_{t}]-X_{t}^{(i)}}{X_{t}^{(i)}} = r_t^{(i)},
\ee
where $r_t^{(i)}$ is the return of the asset $i$ between $t$ and $t+1$. As
previously, we will assume in what follows that $r_t^{(i)}= r^{(i)}$
is time independent. Thus, the rational expectation condition for the
assets
$i$ reads
\ba
X_t^{(i)} &=& \delta^{(i)} \cdot E_{{\mathbb P}_X}\left[ X^{(i)}_{t+1}
| {\cal F}_t\right]~,
\\ &=&  \delta^{(i)} \cdot E_{\mathbb P}\left[ X_{t+1}^{(i)} |
{\cal F}_t\right]~,
\ea
where $\delta^{(i)}$ is the discount factor.

A priori, each asset has a different return. Thus, introducing the
vector {\bf \~X}$_t$ whose
$i^{th}$ componant is $\frac{X_t^{(i)} }{\delta^{(i)}}$, we can
summarize the rational
expectation condition as
\be
\label{eq:rec}
{\bf \tilde X_t} = E_{\mathbb P}\left[ {\bf X_{t+1}} | {\cal F}_t\right].
\ee

Again, we evaluate the conditional expectation of (\ref{eq:sre}), and
using the fact
that $\{ {\bf A}_t\}$ are i.i.d., we have
\be
E_{\mathbb P}\left[ {\bf X_{t}} | {\cal F}_{t-1}\right] = E_{\mathbb
P}[{\bf A}]
{\bf X_{t-1}}.
\ee
This equation together with (\ref{eq:rec}), leads to
\be
{\bf \tilde X_{t-1}} =E_{\mathbb P}[{\bf A}] {\bf X_{t-1}},
\ee
which can be rewritten
\be
\label{eq:45}
\left(E_{\mathbb P}[{\bf A}]-{\bf \widetilde{\delta^{-1}}}
  \right) {\bf X_{t-1}}=0,
\ee
where ${\bf \widetilde{\delta^{-1}}}=
\mbox{diag}[{\delta^{(1)}}^{-1}\cdots
  {\delta^{(d)}}^{-1}]$
  is the matrix whose $i^{th}$ diagonal component
is ${\delta^{(i)}}^{-1}$ and $0$ elsewhere.

The equation (\ref{eq:45}) must be true for every ${\bf X_{t-1}} \in
{\mathbb R}^d$, thus
\be
E_{\mathbb P}[{\bf A}]={\bf \widetilde{\delta^{-1}}}~,
\ee
which is the result announced in section (\ref{sec:nflrs}).

\section*{APPENDIX C: proof of Proposition 2 on the condition
$\kappa_1<1$}

{\it First step} : Behavior of the function
\be
f(\kappa)=\lim_{n \rightarrow \infty}
   {1 \over n} \ln E_{{\mathbb P}_A}\left[ ||{\bf A}_n \cdots {\bf
A}_1||^{\kappa_1}
   \right]
   \label{akdala}
   \ee
   in the interval $[0,\kappa_0]$.

In \cite{K73}, Kesten shows that the function $f$ has the following
properties :
\begin{itemize}
\item $f$ is continuous on $[0,\kappa_0]$,
\item $f(0)=0$ and $f(\kappa_0)>0$,
\item $f'(0)<0$ (this results from the stationarity condition),
\item $f$ is convex on $[0,\kappa_0]$.
\end{itemize}

Thus, there is a unique solution $\kappa_1$ in $(0,\kappa_0)$ such that
$f(\kappa_1)=0$. In order to have $\kappa_1 >1$, it is necessary that
$f(1)<0$, or using the definition of $f$:
\be
\label{eq:cn}
\lim_{n \rightarrow \infty} {1 \over n} \ln E_{{\mathbb P}_A}
||{\bf A}_n \cdots {\bf A}_1|| < 0.
\ee

\begin{figure}
\begin{center}
\epsfig{file=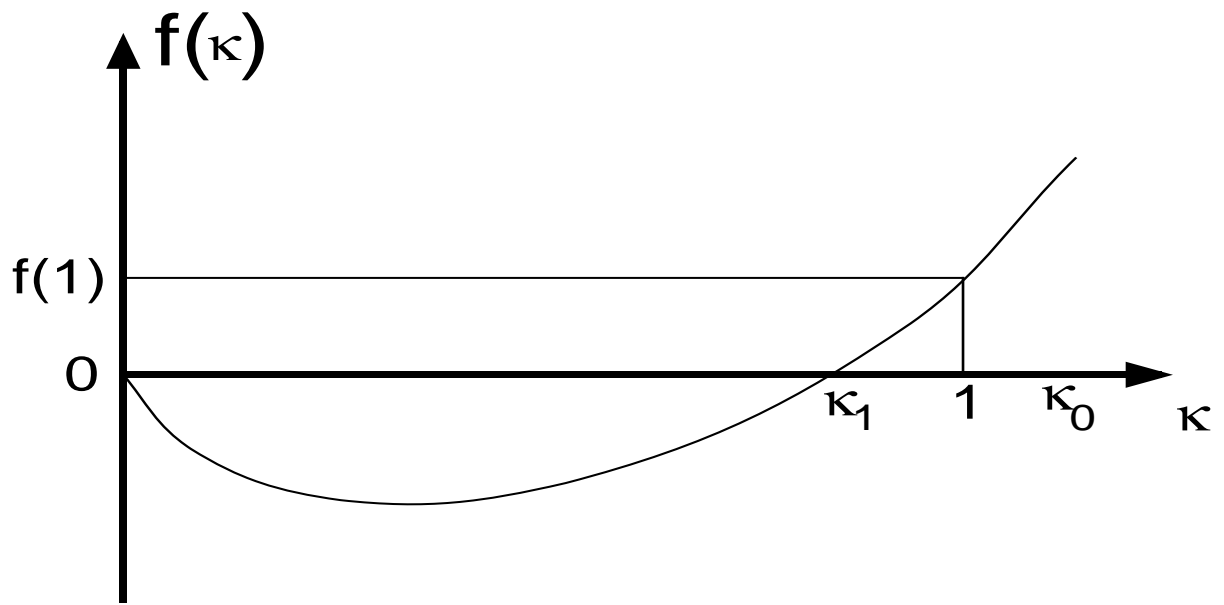,height=8cm,width=16cm}
\caption{\protect\label{fkap} Schematic shape of the function $f(\kappa)$
defined in (\ref{akdala}).
}
\end{center}
\end{figure}

The qualitative shape of the function $f(\kappa)$ is shown in figure
\ref{fkap}.

\vspace{1cm}

\noindent
{\it Second step}~:

  The operator $|| \cdot ||$ is convex:
\be
\forall \alpha \in [0,1]~~\mbox{and}~~\forall ({\bf A},{\bf C})~d \times d
\mbox{-matrices,}~~|| \alpha {\bf A} + (1-\alpha) {\bf C} || \le \alpha
||{\bf A}||  + (1-\alpha) ||{\bf C}||~.
\ee
Thanks to Jensen's inequality, we have
\be
E_{{\mathbb P}_A}||{\bf A}_n \cdots {\bf A}_1|| \ge ||E_{{\mathbb P}_A}
[{\bf A}_n \cdots {\bf A}_1]||~.
\ee
The matrices $({\bf A}_n)$ being i.i.d., we obtain
\be
E_{{\mathbb P}_A}||{\bf A}_n \cdots {\bf A}_1|| \ge || \left\{E_{{\mathbb
P}_A}
[{\bf A}] \right\}^n||~.
\ee

Now, let consider the normalized-eigenvector $x_{\max}$ associated with
the
largest eigenvalue
\be
\lambda_{\max} \equiv \rho(E_{{\mathbb P}_A}[{\bf A}])~,
\ee
where $\rho(E_{{\mathbb P}_A}[{\bf A}])$ is the spectral radius of ${\bf
A}$.
By definition,
\be
|| \left\{ E_{{\mathbb P}_A}[{\bf A}] \right\}^n|| \ge
| \left\{ E_{{\mathbb P}_A}[{\bf A}] \right\}^n x_{\max}| =
   \lambda_{\max}^n  ~.
\ee
Then
\be
\lim_{n \rightarrow \infty} {1 \over n} \ln E_{{\mathbb P}_A}||{\bf A}_n
\cdots
{\bf A}_1|| \ge
\lim_{n \rightarrow \infty} {1 \over n} \ln \rho(E_{{\mathbb
P}_A}[{\bf A}])^n =
   \ln \rho(E_{{\mathbb P}_A}[{\bf A}])~.
\ee

Now, suppose that $\rho(E_{{\mathbb P}_A}[{\bf A}]) \ge 1$. We obtain
\be
f(1)=\lim_{n \rightarrow \infty} {1 \over n} \ln E_{{\mathbb
P}_A}||{\bf A}_n \cdots
   {\bf A}_1|| \ge 0,
\ee
which is in contradiction with the necessary condition (\ref{eq:cn}).

Thus,
\be
f(1)< 0 \Longrightarrow  \rho(E_{{\mathbb P}_A}[{\bf A}])<1.
\ee

\pagebreak

\end{document}